# First-Principles Modeling of Equilibration Dynamics of Hyperthermal Products of Surface Reactions Using Scalable Neural Network Potential

Qidong Lin[1] and Bin Jiang[1,2*]

*1. Key Laboratory of Precision and Intelligent Chemistry, Key Laboratory of Surface and Interface Chemistry and Energy Catalysis of Anhui Higher Education Institutes, Department of Chemical Physics, University of Science and Technology of China, Hefei, Anhui 230026, China*

*2. Hefei National Laboratory, University of Science and Technology of China, Hefei, 230088, China*

Equilibration dynamics of hot oxygen atoms following $O_2$ dissociation on Pd(100) and Pd(111) surfaces are investigated by molecular dynamics simulations based on a scalable neural network potential enabling first-principles description of $O_2$ and O interacting with variable Pd supercells. We find that to accurately describe the equilibration dynamics after dissociation, the simulation cell length necessarily exceeds twice the maximum distance of equilibrated oxygen adsorbates. By analyzing hundreds of trajectories with appropriate initial sampling, the measured distance distribution of equilibrated atom pairs on Pd(111) is well reproduced. However, our results on Pd(100) suggest that the ballistic motion of hot atoms predicted previously is a rare event under ideal conditions, while initial molecular orientation and surface thermal fluctuation could significantly affect the overall post-dissociation dynamics. On both surfaces, dissociated oxygen atoms remain primarily locate their nascent positions and then randomly cross bridge sites nearby.

Elementary surface reactions are vital cornerstones of a variety of interfacial chemical processes such as heterogeneous catalysis and crystal growth. It is commonly assumed that the energy release in each elementary step is almost instantaneously equilibrated by the substrate serving like a bath. This presumption allows Markovian treatments of individual decoupled thermal reactions in the microkinetic modeling of overall processes[1,2]. However, for a reaction with strong exothermicity, the releasing chemical energy is appreciable that unlikely dissipates into the substrate immediately after reaction. The nascent products can be thus highly mobile and active, acting as so-called "hot precursors"[3] that can influence subsequent reaction and diffusion steps in practical surface processes[4].

One of such prototypes is the dissociative adsorption of molecular oxygen on metal surfaces[5-9]. Either activated or non-activated, the dissociation of $O_2$ on different metal surfaces generally releases a large amount of energy (2~3 eV)[10], triggering hyperthermal travel of product oxygen atoms across multiple sites. By suppressing thermal diffusion at sufficiently low temperatures, scanning tunneling microscopy (STM) experiments have identified the separation of dissociated oxygen adatom pairs after equilibration, which ranged from one to more than a dozen of surface lattice constants (SLCs) on various surfaces[5-9]. Unfortunately, these experiments detected the equilibrated product state only and provided no microscopic insights into the equilibration dynamics. This highlights the need for detailed dynamical simulations to understand these experimental findings.

Modelling the equilibration of hyperthermal adsorbates is practically more challenging than typical direct gas-surface scattering processes[11], where incident molecules only interact with a localized surface region for a limited time. While in the former the surface model should be large enough to cover the long distance of adsorbate traveling and the propagation of excited phonons[12-19]. Early such simulations were largely based on empirical potentials[12] or generalized Langevin oscillator models[14]. More recently, ab initio molecular dynamics (AIMD)[13,20] at the density-functional theory (DFT) level were conducted within a confined supercell. To overcome the size limit, Reuter and coworkers proposed an elegant hybrid quantum-mechanics/metal embedding (QM/Me) model[16,18,19], in which the reaction center for molecular dissociation described by AIMD is embedded into an extended metallic environment characterized by an empirical potential. With this QM/Me-AIMD scheme, they revealed much faster heat dissipation from hot oxygen atoms to surface phonons on Pd(100) than Pd(111). The former was attributed to a ballistic type of motion and the latter a random-walk-type diffusion that leads to a smaller net atom pair separation[18]. Nevertheless, the high computational cost of AIMD prevents these QM/Me studies from quantitatively predicting the final product distance distribution and from analyzing its dependence on different initial conditions and supercell sizes. In addition, the QM/Me model relies on an ansatz separating the chemical and elastic contributions in the QM treatment[16], whose reliability is likely system-dependent.

In this Letter, we propose a more efficient way to attack this problem using a global and scalable neural network potential energy surface (PES) for the molecule-surface entirety describing both molecular dissociation and hot atom diffusion. Atomistic neural network PESs developed in recent years[21-23] have been applied to molecular adsorption/desorption in small supercells[24-28], allowing MD simulations as accurate as AIMD but orders of magnitude faster. In particular, the atomistic form enables the additivity of atomic energies and the scalability of the PES when increasing the size of supercell[29,30], without sacrificing the level of interatomic description of the bath as in the QM/Me model.

To validate our approach, we revisit the energy dissipation dynamics of hyperthermal oxygen atoms after $O_2$ dissociation on Pd(100) and Pd(111) that have attracted great experimental[9] and theoretical interests[16-19,31]. To this end, a unified PES for the $O_2$/Pd system was constructed by the embedded atom neural network (EANN) approach[32,33]. The total energy is given by a sum of atomic energies and each atomic energy is an output of an atomic neural network whose input is an array of embedded atom density-like descriptors that are optimized to characterize the atom-centered local environment. Training data were computed by the spin-polarized DFT implemented in the Vienna ab initio simulation package (VASP)[34,35] using the Perdew-Burke-Ernzerhof (PBE) functional[36]. To ensure the scalability of the PES, the training set includes data points not only from four-layer slabs within a $3 \times 3$ supercell (labelled as $3 \times 3 \times 4L$ hereafter) for describing $O_2$ dissociation on both surfaces, but also from larger slabs, $5 \times 5 \times 8L$ and $5 \times 5 \times 7L$ for describing oxygen diffusion on Pd(100) and Pd(111), respectively. A total of 7925 reference points were collected by an uncertainty-driven active learning scheme[37] to yield an analytical EANN PES. MD simulations were performed with different initial samplings described below and a maximum trajectory time of 8 ps. More computational details are given in the Supplemental Material (SM)[38].

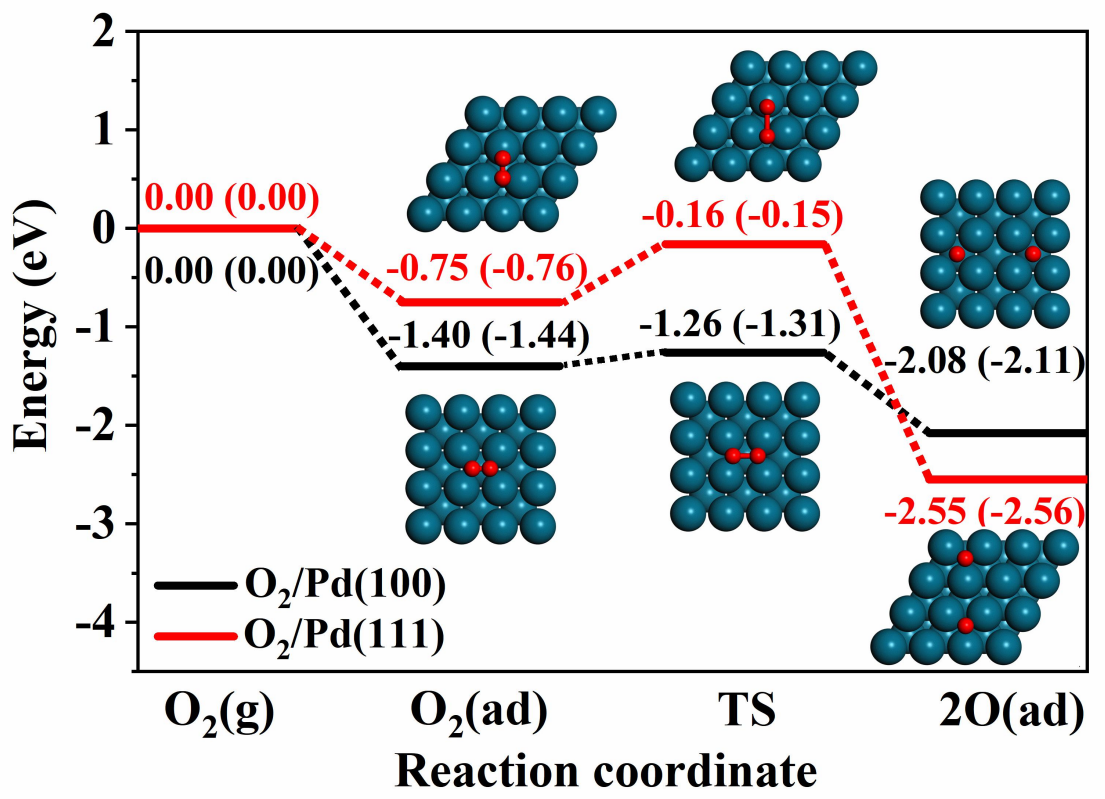

FIG. 1. Minimum energy paths of $O_2$ dissociation on Pd(100) and Pd(111) obtained from the EANN PES and DFT (in parentheses) calculations.

The overall quality of the EANN PES is quite good, evidenced by the fitting root-mean-square-error (RMSE) of 0.3 meV/atom for energy and 39.9 meV/Å for atomic forces. The PES also well depicts the adsorption and dissociation minimum energy paths on both surfaces, as compared with DFT results in Fig. 1. Specifically, $O_2$ first adsorbs molecularly and then overcomes a transition state (TS) for the dissociative adsorption, giving rise to a comparably large exothermicity over 2 eV. Interestingly, the molecular adsorption well depth on Pd(100) is about twice that on Pd(111) (1.40 eV *vs* 0.75 eV), whereas the dissociation barrier height relative to the adsorption well is in turn much lower on Pd(100) than on Pd(111) (0.14 eV *vs* 0.59 eV). Consequently, the initial adsorption energy on Pd(100) is large enough to proceed to direct dissociation[17], while on Pd(111) the molecularly adsorbed $O_2$(ad) precursor needs sufficient heating to overcome the dissociation barrier[39]. We therefore adopt different initial samplings in subsequent MD simulations. In addition, the diffusion barrier of an oxygen adatom nearby a co-adsorbed oxygen on Pd(100) is ~0.23 eV (0.25 eV by DFT), somewhat lower than that ~0.33 eV (0.34 eV by DFT) on Pd(111). These numbers agree reasonably well with previous predictions[19]. Additional results in Figs. S1-S4 further validate the reliability and scalability of the PES.

Let us now discuss the equilibration dynamics of hyperthermal oxygen adatoms on Pd(100). Meyer and Reuter[16] showed that a $3 \times 8 \times 3L$ slab for Pd(100) used in stand-alone AIMD simulations failed to predict the equilibrium separation of O adatoms. It is therefore interesting to investigate the dependence of the equilibration dynamics on the cell size. To this end, we keep the initial condition of the exemplary trajectory close to that in Ref. [19], in which the $O_2$ molecule lies side-on at 1.8 Å above the hollow site along the [001] axis with a negligible kinetic energy of 25 meV normal to an equilibrium surface of 0 K. Fig. 2 shows the evolution of the total kinetic energy of two oxygen adsorbates ($E_{k\text{-O}}$) and the O-O distance ($d_{\text{O-O}}$) along the trajectory in different supercells. In all cases, the exothermic dissociation of the parent $O_2$ molecule create hyperthermal oxygen products and leads them to locate neighboring hollow sites separated by ~2 SLCs of Pd(100) (~5.6 Å) in the first ~0.25 ps. However, the smallest $3 \times 3$ supercell cannot facilitate further separation of oxygen atoms due to the presence of artificial repulsion caused by oxygen atoms within image cells. Additionally, part of the energy transferred to surface atoms is reflected by the cell boundary and directed back towards the adsorbates, leading to a pronounced oscillation of $E_{k\text{-O}}$. Increasing the cell size enlarges the bath capacity and expands the surface boundary, thus weakening energy reflection and fluctuation and allowing for larger $d_{\text{O-O}}$ values. Accordingly, oxygen adatoms dissipate their kinetic energy to the lattice as they further depart up to ~0.6 ps, reaching a maximum separation of 4 SLCs. While the O-O distance exhibits a decrease in smaller supercells due to the repulsion between image atoms, it stabilizes and converges at a cell size of $8 \times 8$, whose length (~8 SLCs) is twice the final $d_{\text{O-O}}$ and sufficient to avoid artificial image interactions. We note that the energy dissipation in the perpendicular direction quickly converges with ~2 moving Pd layers (see Fig. S5).

Interestingly, we find that an asymmetric $3 \times 8$ supercell can also result in a large $d_{\text{O-O}}$ of ~4 SLCs, as shown in Fig. 2b. This is possible if the ballistic-type motion of hot oxygen atoms on Pd(100) along a hollow-bridge-hollow path[18] parallel to the longer length (8 SLC) in this cell.

In contrast, the reason for the failure of the AIMD work of Meyer and Reuter[16] using the same supercell and density functional is less clear. One possibility is that oxygen atoms in their calculation happen to diffuse along the shorter side of the cell, which is too short to prevent the repulsion between image adsorbates (their result is similar to our $3 \times 3$ result here). To further monitor the actual energy flow in the substrate, we analyze in Fig. 2c the kinetic energy of all Pd atoms and that of those in a $3 \times 8 \times 4L$ reaction center embedded into a $10 \times 10 \times 10L$ slab. In good agreement with the QM/Me work[16], the released chemical energy quickly propagates out of the embedding region to the entire slab, which eventually takes up ~75% of the initial energy at 1.5~2.0 ps. This result suggests that the $10 \times 10 \times 10L$ slab used here works similarly well as the much larger $50 \times 50 \times 50L$ phononic bath in the QM/Me model[16]. Even certain energy reflection at the cell boundary may occur, it is unlikely to impact the diffusion dynamics of the oxygen adatoms in this timescale.

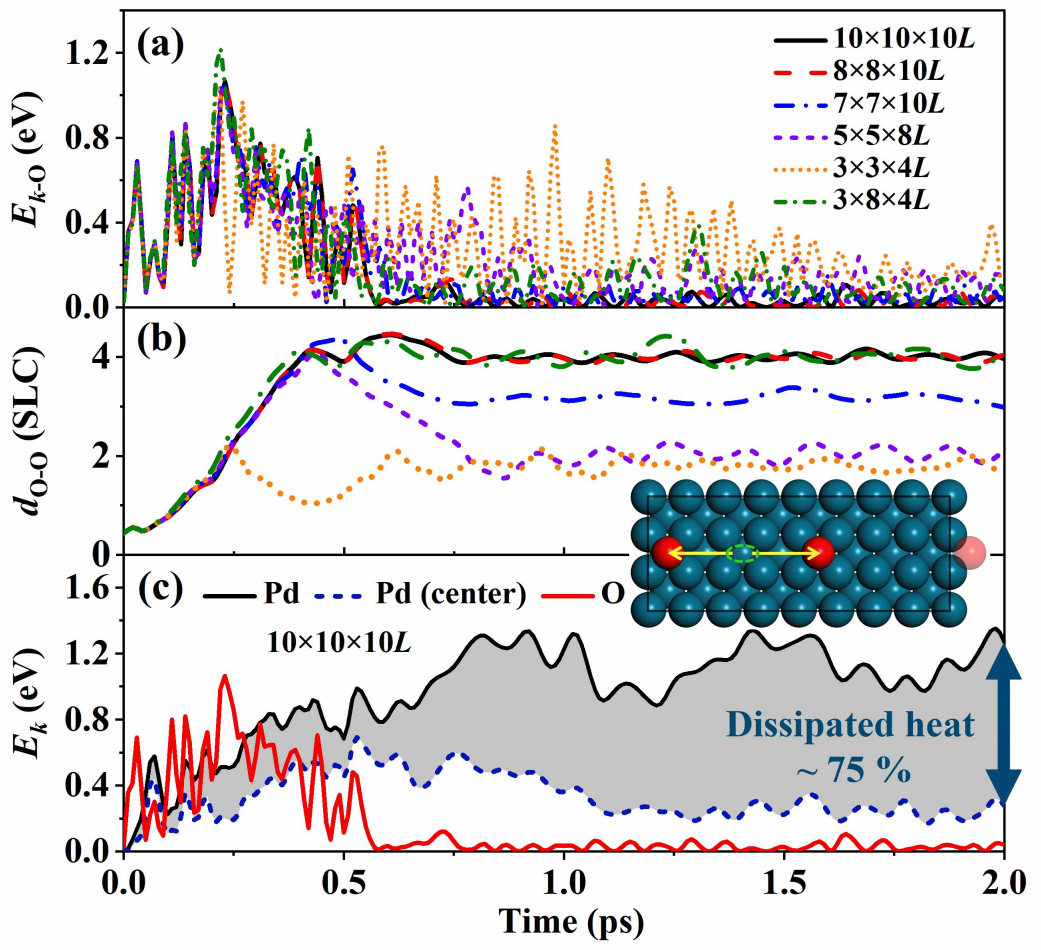

FIG. 2. (a) Kinetic energy ($E_{k\text{-O}}$) of and (b) distance between oxygen atoms ($d_{\text{O-O}}$ in SLC) as a function of time along representative trajectories of $O_2$ dissociation on Pd(100) with different supercell sizes. (c) Kinetic energy of oxygen adsorbates (red line), all Pd atoms (black line) of a $10 \times 10 \times 10L$ slab, and of those Pd atoms in a $3 \times 8 \times 4L$ local center (dashed blue line) as a function of time. The gray region is a measure for the dissipated heat to surface atoms outside the local center. The inset shows the diffusion path and final oxygen positions, where the transparent red ball schematically illustrates an image oxygen atom in the neighboring supercell.

Similar tests on cell size have been conducted for the $O_2$/Pd(111) system, where $O_2$ is initially placed at the TS with a negligibly small energy imparted along the O-O bond. This allows the precursor-mediated dissociation to occur naturally, driven by the large exothermicity, as demonstrated in Ref. [19]. Fig. S6 shows that $E_{k\text{-O}}$ on Pd(111) is more oscillating than on Pd(100) with fluctuations as large as ~0.5 eV even after 2 ps for the largest $10 \times 10$ supercell tested. The equilibrium $d_{\text{O-O}}$ largely converges at a $7 \times 7$ supercell ($d_{\text{O-O}} = \sqrt{7}$ SLCs of Pd(111), ~7.4 Å). This slow equilibration on Pd(111) is consistent with the QM/Me prediction, which was attributed to the long-lived Rayleigh phonon mode excitations that confine the released energy within a “hot spot” around the reaction zone.[18,19].

Since the EANN PES can describe the large energy dissipation and long product separation on both surfaces with much higher efficiency than the QM/Me-AIMD model, we can now evaluate the equilibrium $d_{\text{O-O}}$ distribution by counting hundreds of trajectories with realistic initial conditions. While there was no STM experiment on Pd(100), Rose *et al.* did observe a broad $d_{\text{O-O}}$ distribution for $O_2$ dissociation on Pd(111) in the range of $1 \sim \sqrt{7}$ SLCs with most probable separations at $\sqrt{3}$ and 2 SLCs, via both thermal dissociation at 160 K and tip-induce dissociation at 40 K[9]. To mimic the direct dissociation on Pd(100), $O_2$ molecules were placed at 5 Å above the surface of 160 K with initial orientations and positions randomized within a unit cell. On the other hand, to ensure the precursor-mediated dissociation on Pd(111), $O_2$ molecules were assumed to locate at the TS above the hcp site of the surface at 160 K with a negligible kinetic energy of 5~50 meV randomly added to each molecular degree of freedom. To be consistent and ensure convergence, a $10 \times 10 \times 10L$ slab is used for both surfaces.

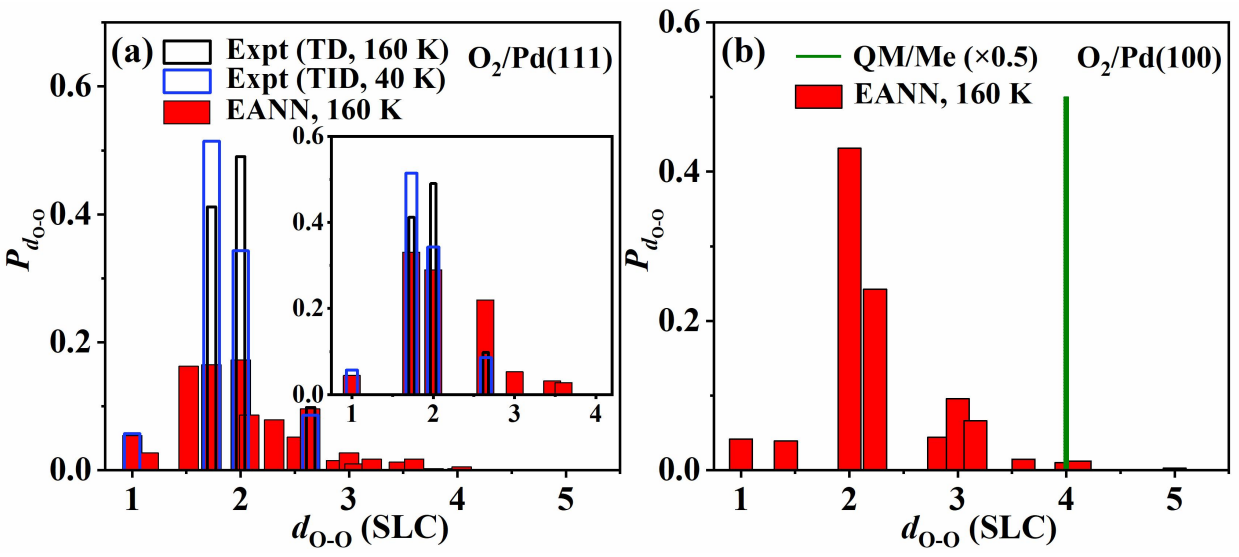

FIG. 3. Comparison of calculated and experimental (whenever available) distance distributions of oxygen atom pairs after $O_2$ dissociation on (a) Pd(111) and (b) Pd(100). See text for details.

Fig. 3 presents the $d_{\text{O-O}}$ distributions over 400 trajectories on Pd(100) and on Pd(111), respectively. Encouragingly, the calculated distribution on Pd(111) reproduces the experimental counterpart reasonably well, although there are more possible $d_{\text{O-O}}$ values in simulations as oxygen atoms can adsorb at the fcc or the hcp site. Provided that oxygen adsorbates at metastable hcp sites will eventually move to nearest fcc sites in a macroscopic timescale, the theory-experiment agreement is much improved, as shown in the inset of Fig. 3a. Note that initializing trajectories at the TS above the fcc site will not change this conclusion (see Fig. S7). Previous QM/Me results for Pd(111) also qualitatively predicted $d_{\text{O-O}}$ in the range of 1~3 SLCs of which the most common is 2 SLCs[19], which were interpreted by a random-walk mechanism[19]. As illustrated in Figs. 4a-d, the TS above the hcp site favors a “top-hcp-bridge” orientation and dissociating oxygen atoms fly apart oppositely along this path. One of the O atoms can bypass the highly energetic top site to a fcc or hcp site in a neighboring cell, while the other passes through the bridge site to the fcc site in the primordial cell, corresponding to $d_{\text{O-O}}$ values of $\sqrt{7/3}$ or $\sqrt{3}$ SLCs, respectively. After that, these adatoms migrate randomly to other hollow sites leading to other $d_{O\text{-}O}$ values (see Fig. S8).

Interestingly, the $d_{O\text{-}O}$ distribution on Pd(100) here is also found to cover a wide range of distances between four-fold hollow sites within 1~5 SLCs with dominant population at 2 SLCs. This result is in striking contrast to a single $d_{O\text{-}O}$ value (~4 SLCs) predicted by nine representative QM/Me trajectories[16,18,19]. In QM/Me trajectories (and the exemplary trajectory displayed above in Fig. 2), the parent $O_2$ molecule was perfectly aligned towards the "bridge-hollow-bridge" direction parallel to the [001] axis above an equilibrium surface at 0 K. The molecule straightly dissociates and the atomic products immediately separate along this TS orientation, as illustrated in Fig. S9. Unlike the case on Pd(111) where oxygen atoms have to detour the highly energetic top site, this dissociation path on Pd(100) passes through the low energetic bridge site with minimal hindrance. Moreover, oxygen atoms can push away the top-layer Pd atoms symmetrically when overcoming the diffusion barrier and eventually stop by ~4 SLCs after running out of their kinetic energy.

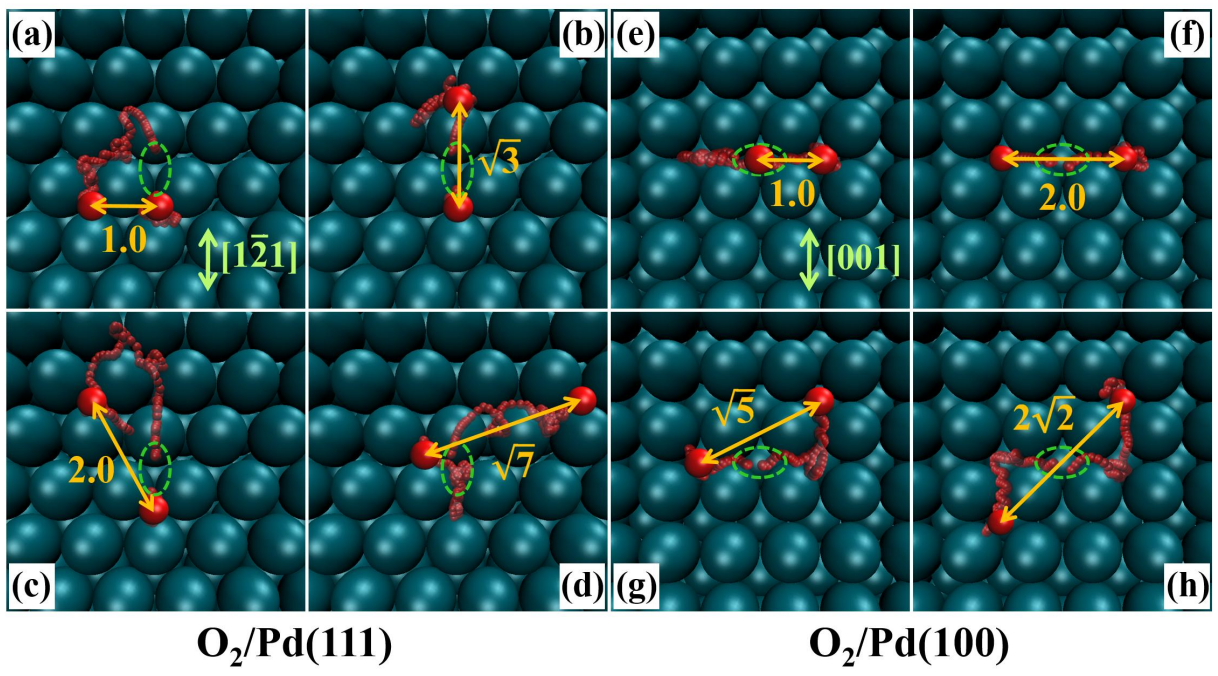

FIG. 4. Evolution of atomic oxygen positions during representative trajectories of post-dissociation dynamics of $O_2$ on (a-d) Pd(111) and (e-h) Pd(100) at 160 K, leading to different equilibrated distances (orange numbers in SLC). The dissociation region is circled in green and oxygen atoms evolve as red balls (Final positions are enlarged for visibility).

However, with the more realistic sampling of molecules and surface configurations at 160 K, molecules with imperfect positions and orientations will steer to the favorable TS for dissociation, but leaving some residual momenta in other directions. Meanwhile, the thermal fluctuation of Pd atoms will increase the diffusion barrier and provide asymmetric hindrance to the oxygen diffusion. As a result, the ballistic travel of the hyperthermal oxygen atom along a straight line is rarely seen. In practice, we find in most trajectories that these oxygen atoms can easily move outside the reaction center to the next hollow sites along the [001] direction separated by 2 SLCs. More than 40% of them stabilize this status with energy dissipation to the substrate by collisions with surrounding Pd atoms. Other ones can move either forward, or perpendicularly, or even backward, as shown in Fig. 4e-h and Fig. S8, resulting in multiple $d_{O\text{-}O}$ values. Only ~1% of them reach 4 SCLs as an occasional result of extra forward hops of one of oxygen adsorbates. Our results suggest a similar random-walk-type diffusion of hot O atoms on Pd(100) as on Pd(111), if realistic conditions are considered. The difference is that each hop leads a smaller net displacement of $d_{O\text{-}O}$ on Pd(111) than on Pd(100) due to their different surface symmetry. We note that our conclusion would not be significantly altered by electron-hole pair excitations. Combining a widely-used local density friction approximation (LDFA)[40] with an EANN fitted electron density surface (detailed in the SM), we find that this nonadiabatic channel barely changes the energy loss profile of the equilibration process, contributing ~10% of total chemisorption energy loss and occurring largely after oxygen atoms are already trapped (Fig. S10). This confirms the dominant role of phonon-mediated energy dissipation in thermalizing hot O atoms on Pd surfaces[18,31].

To summarize, we study the equilibration dynamics following the exothermic $O_2$ dissociation on Pd(100) and Pd(111) based on a scalable neural network potential that describes a varying size of supercells with first-principles accuracy. We clarify that the simulation cell length necessitates twice the maximum separation of thermalized adsorbates. More importantly, our results indicate that the previously recognized ballistic movement of hot O atoms on Pd(100) was an accidental result of an ideal initial molecular orientation and surface configuration[16]. With more realistic sampling in many trajectories, we discover that on both surfaces, hot oxygen atoms tend to locate most favorable neighbor sites after dissociation, followed by a similar random-walk-type motion across bridge sites, although on Pd(111) residual kinetic energy is higher. This mechanism leads to a finite distance distribution of equilibrated atom pairs consistent with experiment on Pd(111). Thanks to its high efficiency and scalability, this neural network approach applied here is generally suitable for future study of dynamical processes occurring in even larger surface areas and various initial conditions[8,41].

This work is supported by the Strategic Priority Research Program of the Chinese Academy of Sciences (XDB0450101), Innovation Program for Quantum Science and Technology (2021ZD0303301), CAS Project for Young Scientists in Basic Research (YSBR-005), National Natural Science Foundation of China (22221003, 22073089, and 22033007). We acknowledge the Supercomputing Center of USTC, Hefei Advanced Computing Center, Beijing PARATERA Tech CO., Ltd for providing high performance computing service.

*bjiangch@ustc.edu.cn

# Supplemental Material

## First-Principles Modeling of Equilibration Dynamics of Hyperthermal Products of Surface Reactions Using Scalable Neural Network Potential

Qidong Lin[1] and Bin Jiang[1,2*]

*1. Key Laboratory of Precision and Intelligent Chemistry, Key Laboratory of Surface and Interface Chemistry and Energy Catalysis of Anhui Higher Education Institutes, Department of Chemical Physics, University of Science and Technology of China, Hefei, Anhui 230026, China*

*2. Hefei National Laboratory, University of Science and Technology of China, Hefei, 230088, China*

*: corresponding author: bjiangch@ustc.edu.cn

# 1. Computational Details

## 1.1 Density Functional Theory

In this work, all energies and atomic forces used for fitting the neural network (NN) potential energy surface (PES) were obtained through spin-polarized plane-wave density functional theory (DFT) calculations using the Vienna ab initio simulation package (VASP) code[1, 2] and the generalized gradient approximation (GGA) based Perdew-Burke-Ernzerhof (PBE) functional[3]. The projector-augmented wave (PAW) method[4, 5] was used to describe the interaction between ionic cores and electrons. The kinetic energy of the plane-wave basis set was truncated at 400 eV. First, we modeled Pd(100) and Pd(111) with a four-layer slab in a 3×3 supercell with the top two layers relaxed (labeled as a 3×3×4*L*(2*R*) slab model hereafter) to describe $O_2$ dissociation, for which a 4×4×1 Gamma-centered *k*-point mesh was used to sample the Brillouin zone. To describe the hyperthermal motion of oxygen adatoms, additional large-sized 5×5×8*L*(7*R*) and 5×5×7*L*(6*R*) slabs for Pd(100) and Pd(111) were included with a single Gamma-centered *k*-point. To evaluate the nonadiabatic energy dissipation channel, one needs to learn the spatial distribution of the electron density of a clean surface. To this end, a 3×8×4*L*(2*R*) slab was modeled for Pd(100) with a 3×1×1 Gamma-centered *k*-point and thermostatic ab initio molecular dynamics (AIMD) simulations of a clean surface were performed with this slab at a surface temperature ($T_s$) of 800 K. All Pd(100) and Pd(111) slabs were separated by a vacuum of 15 Å in the *Z* direction.

## 1.2 Embedded Atom Neural Network Representation

The embedded atom neural network (EANN) method[6] was used to construct a high-dimensional PES that can describe both $O_2$ dissociation and hot atom diffusion on Pd(100) and Pd(111) surfaces. In the EANN framework, the total energy of the system is regarded as the sum of all atomic energies. Each atomic energy is an output of an atomic NN, which is expressed by a virtual embedded electron density at that atomic position, created by neighboring atoms (but not the center)[6],

$$E = \sum_{i=1}^{N} E_i = \sum_{i=1}^{N} \mathrm{NN}_i\left(\boldsymbol{\rho}^i\right). \qquad (1)$$

The embedded atom density-like (EAD) descriptors $\boldsymbol{\rho}^i$ can be constructed by a set of Gaussian-type atomic orbitals (GTO) centered at neighboring atoms $i$, via the square of their linear combination,

$$\rho^i = \sum_{l_x,l_y,l_z}^{l_x+l_y+l_z=L} \frac{L!}{l_x!l_y!l_z!}\left[\sum_{i\neq j}^{N_c} c_j \varphi\left(\mathbf{r}_{ij}\right) f_c\left(r_{ij}\right)\right]^2, \qquad (2)$$

where $N_c$ is the number of atoms near the embedded atom within a given cutoff radius ($r_c$). The cutoff function $f_c(r_{ij})$ ensures that the contribution of each neighboring atom decays smoothly to zero at $r_c$. GTOs can be expressed as

$$\varphi\left(\mathbf{r}_{ij}\right) = x^{l_x} y^{l_y} z^{l_z} \exp\left(-\alpha\left|r_{ij} - r_s\right|^2\right), \qquad (3)$$

where $\mathbf{r}_{ij}$ represents the vector of the embedded atom $i$ relative to atom $j$ in Cartesian coordinates, and $r_{ij}$ is its norm; $l_x$, $l_y$ and $l_z$ are angular momentum projections on each axis, and $L$ is the sum of them; $\alpha$ and $r_s$ are the hyperparameters for determining the radial distribution of GTOs. $c_j$ is the element-dependent expansion coefficient of an atomic orbital for atom $j$, which is also optimized during training. The EANN method incorporates the permutational, translational and rotational symmetry of chemical systems. It is in principle scalable to a large system when being trained well with data points in small systems due to its additivity of atomic energies. It can be also adapted to learn the electron density (as discussed below) and other electronic properties.

An uncertainty-driven active learning scheme[7, 8] was applied to generate a training set for fitting the EANN PES. In this approach, a weighted negative of squared difference surface (NSDS) is defined by two trial NNs with different parameters. The weighted NSDS is closed to zero at existing data locations and has largely negative values in regions that are distant from samples. One searches for the local minima on NSDS corresponding to high uncertainty locations on the PES where ab initio calculations should be performed and added into the training set. This is followed by retraining and the update of trial PESs for a new search. In this iterative way, we started from 4 seed points in a 3×3×4$L$(2$R$) supercell to describe $O_2$ dissociation on Pd(100) and Pd(111), and terminated at 1360 and 3265 data points, respectively. The PES based on these data points were found not able to predict the DFT data for an extended 5×5 supercell or a thicker slab, as discussed below (Figs. S2-S4). Indeed, when training the model with small supercells with a cutoff radius ($r_c$=6.0 Å) exceeding $\alpha_L/2$ ($\alpha_L$ is the simulation box length, *e.g.*, 8.4 Å for the 3×3 supercell and 5.9 Å for the four surface layers of Pd(100)), there are always mirror atoms in the cutoff sphere that move synchronously. Whereas in an extended cell, all atoms in the cutoff sphere will be freely mobile and their atomic environments may not be fully captured by the trained model. To ensure the proper scalability to the PES for describing hyperthermal diffusion of oxygen adatoms, we further sampled data points in larger-sized 5×5×8$L$ and 5×5×7$L$ supercells for Pd(100) and Pd(111) by the same active learning algorithm using two preliminary PESs obtained with the 3×3×4$L$ supercell. In these slabs, the atomic environment was adequately described because $r_c$ is now smaller than $\alpha_L/2$ in any direction. This collection adding 1628 and 1672 data points for Pd(100) and Pd(111), respectively.

The final EANN PES was trained on 7925 data points in total, including potential energies and atomic forces. The ratio for training and validation sets was 90:10. The hyperparameters of GTOs used here were $L$ = 0, 1, 2, $r_c$ = 6.0 Å, $\alpha$ = 0.83 Å$^{-2}$ and $\Delta r_s$ = 0.49 Å (equally spaced), producing 39 EAD features. Each atomic NN has two hidden layers with 50 neurons in each one.

## 1.3 Classical molecular dynamics simulation

To simulate the equilibration dynamics of hyperthermal oxygen adatoms after $O_2$ dissociation on Pd(100) and Pd(111), we carried out classical molecular dynamics (MD) calculations based on the EANN PES using our modified VENUS code[9] for studying molecule-surface reactions[10]. It should be noted that the dissociation of $O_2$ on Pd(100) and Pd(111) proceeds with different mechanisms. As shown in Fig. S1, the dissociation barrier of the adsorbed $O_2$ on Pd(100) is so low that the impinging molecules will directly dissociate after adsorption with high probability. As a result, on Pd(100), the initial $O_2$ molecules were placed 5.0 Å above the metal surface with the orientation and position randomly sampled in a unit cell. A negligible kinetic energy of 25 meV was added to the molecule perpendicular towards the surface. The initial configurations and velocities of the surface atoms were extracted from MD trajectories equilibrated with an Andersen thermostat at an experimental $T_s$ of 160 K. More than 100 trajectories were propagated up to 8.0 ps with a time step of 0.1 fs using the velocity Verlet algorithm. By contrast, the dissociation barrier of the adsorbed $O_2$ on Pd(111) is quite high compared to the adsorption energy so that the impinging molecule will first adsorb molecularly and partially accommodate to the surface, followed by the precursor-mediated dissociation after adjusting its orientation to fit the transition state (TS) and a sufficient heating by the surface. Consequently, the direct collision of $O_2$ on Pd(111) will have a very small dissociation probability. Since our focus is the post-dissociation dynamics, the initial $O_2$ molecule was placed at the TS on the hcp or fcc site, with a negligible kinetic energy of 5~50 meV randomly added to each molecular DOFs, appropriately ensuring the occurrence of $O_2$ dissociation. Other initial conditions are consistent with those of $O_2$/Pd(100).

To evaluate the non-adiabatic contribution of metallic electron-hole pairs (EHPs), we take the $O_2$/Pd system as an example by employing the widely-used electronic friction model[11] based on the local density friction approximation (LDFA)[12]. This model treats non-adiabatic interactions between the adsorbate nuclei and surface EHPs as electronic frictional forces. The nuclear motion[12] can be described by a Langevin equation as follows,

$$m_i \frac{d^2\mathbf{r}_i}{dt^2} = -\frac{\partial V(\mathbf{r}_i,\mathbf{r}_s)}{\partial \mathbf{r}_i} - \eta_i(\mathbf{r}_i)\frac{d\mathbf{r}_i}{dt} + F_{el}(\eta_i, T_{el}), \qquad (6)$$

where $m_i$ is the mass of gas atoms; $\mathbf{r}_i$ and $\mathbf{r}_s$ are the Cartesian coordinates of gas atoms and surface atoms; $V(\mathbf{r}_i, \mathbf{r}_s)$ represents the potential energy; $\eta(\mathbf{r}_i)$ is the friction coefficient of gas atoms, which is determined by the electron density of the metal surface at the position of $\mathbf{r}_i$. $F_{el}$ represents a random force approximated by a Gaussian white noise[11], which is determined by the electron temperature $T_{el}$ (here $T_{el} = T_s$) and the friction coefficient $\eta$.

In this LDFA model, the key to solving the equation of motion is to obtain the electron density of the metal surface at $\mathbf{r}_i$ to calculate the friction coefficient $\eta$. We used an EANN representation to learn the electron density surface (EDS)[6], which deals with a spatial representation of the electron density of a metal surface as an PES of an atom embedded on the surface. We have shown elsewhere[6] that these two representations are essentially the same in their dimension and symmetry. To this end, six hundred distinct configurations were selected from an AIMD trajectory of a clean $3\times8\times4L(2R)$ Pd(100) slab at $T_s$ = 800 K, and corresponding electron densities in three-dimensional space were extracted for fitting the EANN EDS. The hyperparameters of GTOs were set as $L$ = 0, 1, 2, $r_c$ = 4.0 Å, $\alpha$ = 1.3 Å$^{-2}$ and $\Delta r_s$ = 0.39 Å (equally spaced), producing 33 EAD features. There are two hidden layers in each atomic NN with 50 neurons in each one. The root-mean-square-error of this EANN EDS is as low as $4.9\times10^{-4}$ e/Å$^3$ (with the target density ranging from $8.2\times10^{-5}$ e/Å to 0.94 e/Å), which well reproduces the DFT calculations.

## 2. Additional results

In order to validate the quality of the EANN PES, we first plot the site-specific two-dimensional PES cuts for $O_2$ dissociation on Pd(100) and Pd(111) surfaces in Fig. S1. It is shown that the PES is free of any artificial "holes" and coincides well with the DFT energies at these discrete grids that are not included in the training set. Next, as shown in Figs. S2, the energies of these snapshots of classical trajectories at 3×4×4*L* (2*R*) slabs of Pd(100) and Pd(111) for $O_2$ dissociation at 160 K predicted by the EANN PES are in excellent agreement with the recalculated DFT data. Indeed, these trajectories can be also reproduced by an EANN-S PES trained with 4625 data points in 3×3×4*L* supercells only. However, in Fig. S3, the EANN PES also predicts well the energy variation on larger 5×5×8*L* (4*R*) and 5×5×7*L* (4*R*) supercells for Pd(100) and Pd(111), respectively, whereas the EANN-S PES shows significant errors, due to incomplete learning of the atomic environment. In Fig. S4, we demonstrate that the EANN PES is scalable to describe 4×4×4*L* supercells for both Pd(100) and Pd(111), which was not included for training.

Figure S8 shows the evolution of the O-O distance ($d_{O\text{-}O}$) along the representative trajectories for $O_2$ dissociation on Pd(111) and Pd(100) with different equilibrated $d_{O\text{-}O}$ values. On Pd(111), $O_2$ molecules dissociate along the "top-hcp-bridge" path and nascent atomic oxygen products separate by $\sqrt{7/3}$ or $\sqrt{3}$ SLCs. Afterwards, hyperthermal oxygen atoms undergo random hops to neighboring hollow sites and thus end with different final $d_{O\text{-}O}$ values. Each hop corresponds to an oxygen atomic movement by $\sim\sqrt{3}/3$ SLC, which contributes a very small net increase (or decrease) of $d_{O\text{-}O}$ due to the hexagonal surface symmetry. In comparison, on Pd(100), $O_2$ molecules will steer to the favorable TS structure lying towards the "bridge-hollow-bridge" direction for dissociation. Oxygen atoms quickly move outside the reaction center to the next hollow sites along the [001] direction separated by 2 SLCs. Hyperthermal oxygen atoms undergo similar random-walk-type diffusion, but each hop now corresponds to an oxygen atomic movement by ~1 SLC, thus contributing a more prominent net increase (or decrease) of $d_{O\text{-}O}$.

Figure S9 shows several snapshots at different simulation times of the exemplary trajectory of $O_2$ dissociation on Pd(100) ending at $d_{O\text{-}O}$ of four times the surface lattice constant (SLC) (see Fig. 2). In this trajectory, the initial molecular orientation was perfectly aligned towards the "bridge-hollow-bridge" direction parallel to the [001] axis above an equilibrium surface at 0 K so that the $O_2$ dissociation occurs immediately. The yielded hot oxygen atoms push away the top-layer Pd atoms symmetrically when they cross the bridge site and diffuse perfectly along the [001] direction. This can be clearly seen at snapshots of 130 fs (the first time crossing a bridge site) and 300 fs (the second time).

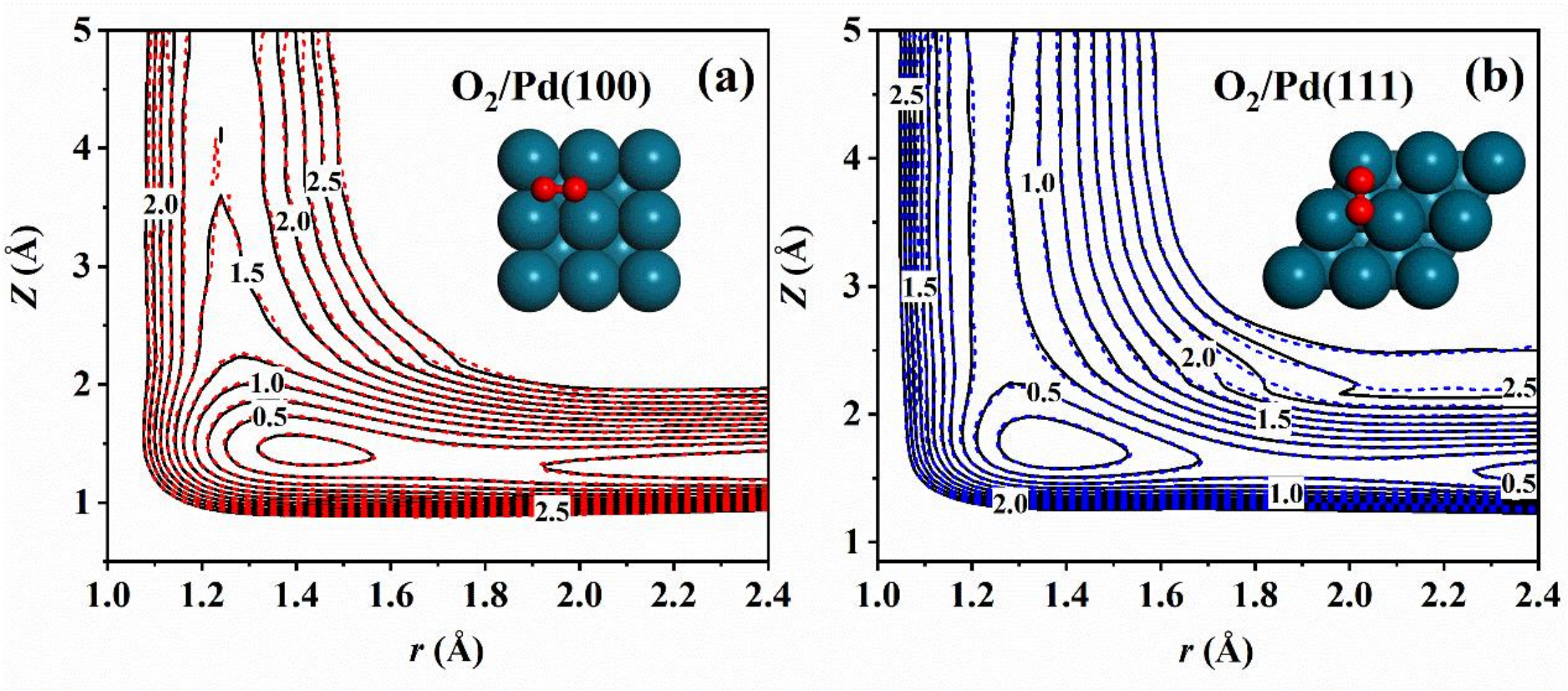

Figure S1. Comparison of two-dimensional contour plots obtained by the EANN PES and direct DFT calculations for (a) $O_2$/Pd(100) and (b) $O_2$/Pd(111) systems, as a function of the bond length of O-O (*r*, in Å) and the molecular height (*Z*, in Å), with the molecular center being fixed at the fourfold hollow site on Pd(100) and threefold hcp site on Pd(111) and the molecular orientation fixed at that of the transition state, respectively, and with other angular coordinates optimized.

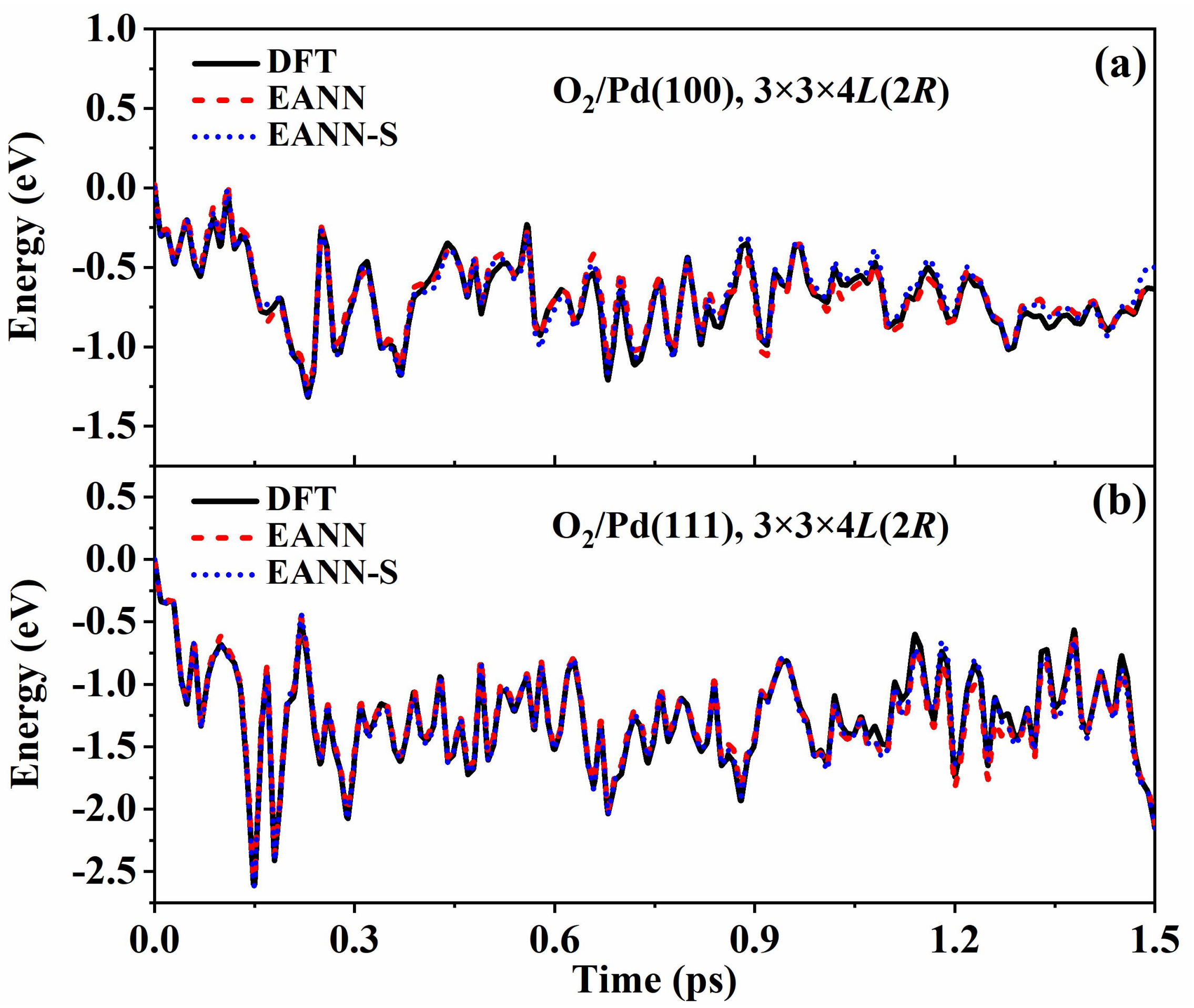

Figure S2. Comparison of the energy variation during a classical trajectory of $O_2$ dissociation on (a) Pd(100) and (b) Pd(111) with 3×3×4$L$(2$R$) supercells at a surface temperature of 160 K obtained from the EANN/EANN-S PESs and direct DFT calculations.

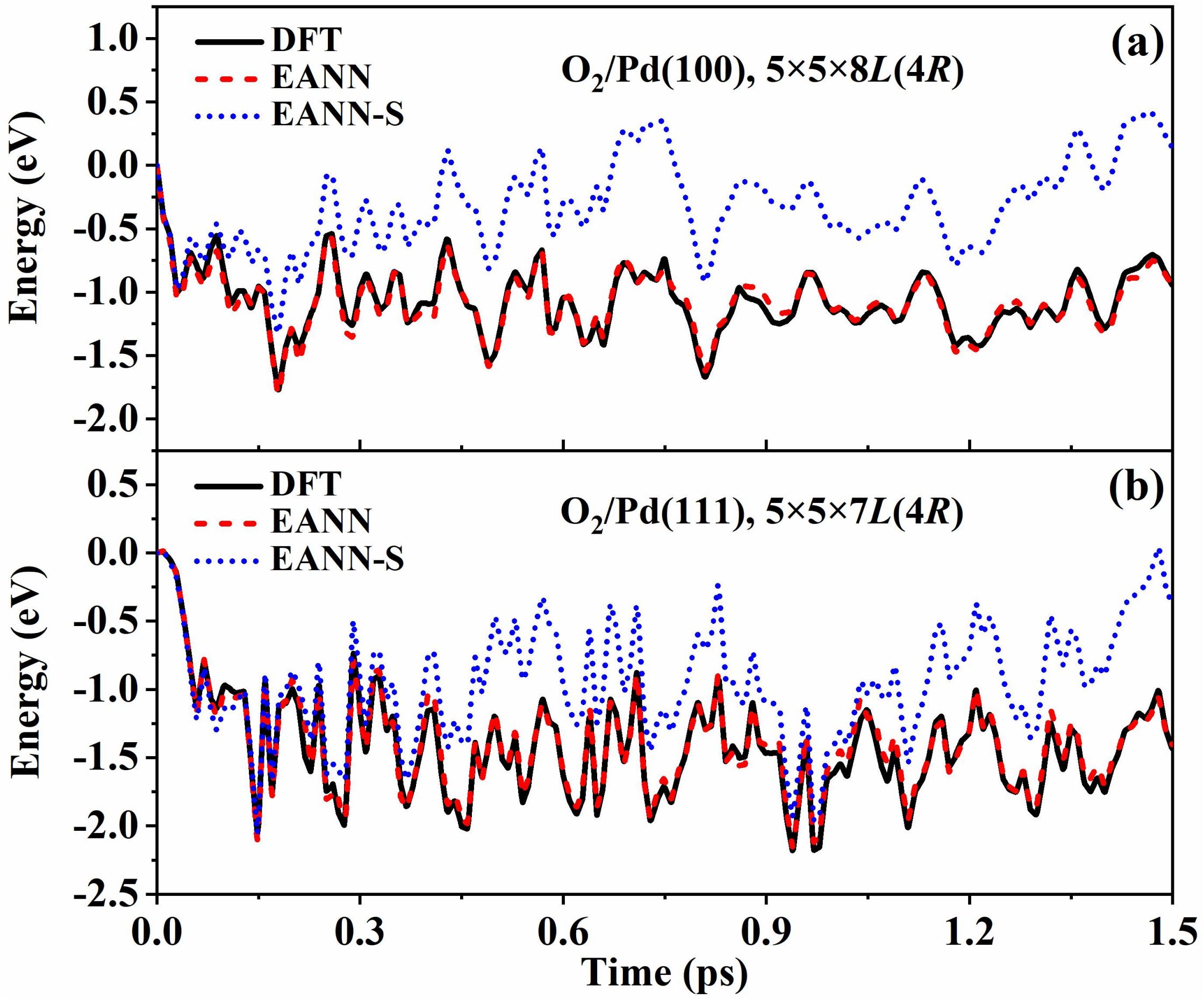

Figure S3. Comparison of the energy variation during a classical trajectory of $O_2$ dissociation on (a) Pd(100) and (b) Pd(111) with 5×5×8*L*(4*R*) and 5×5×7*L*(4*R*) supercells at a surface temperature of 160 K obtained from the EANN/EANN-S PES and direct DFT calculations.

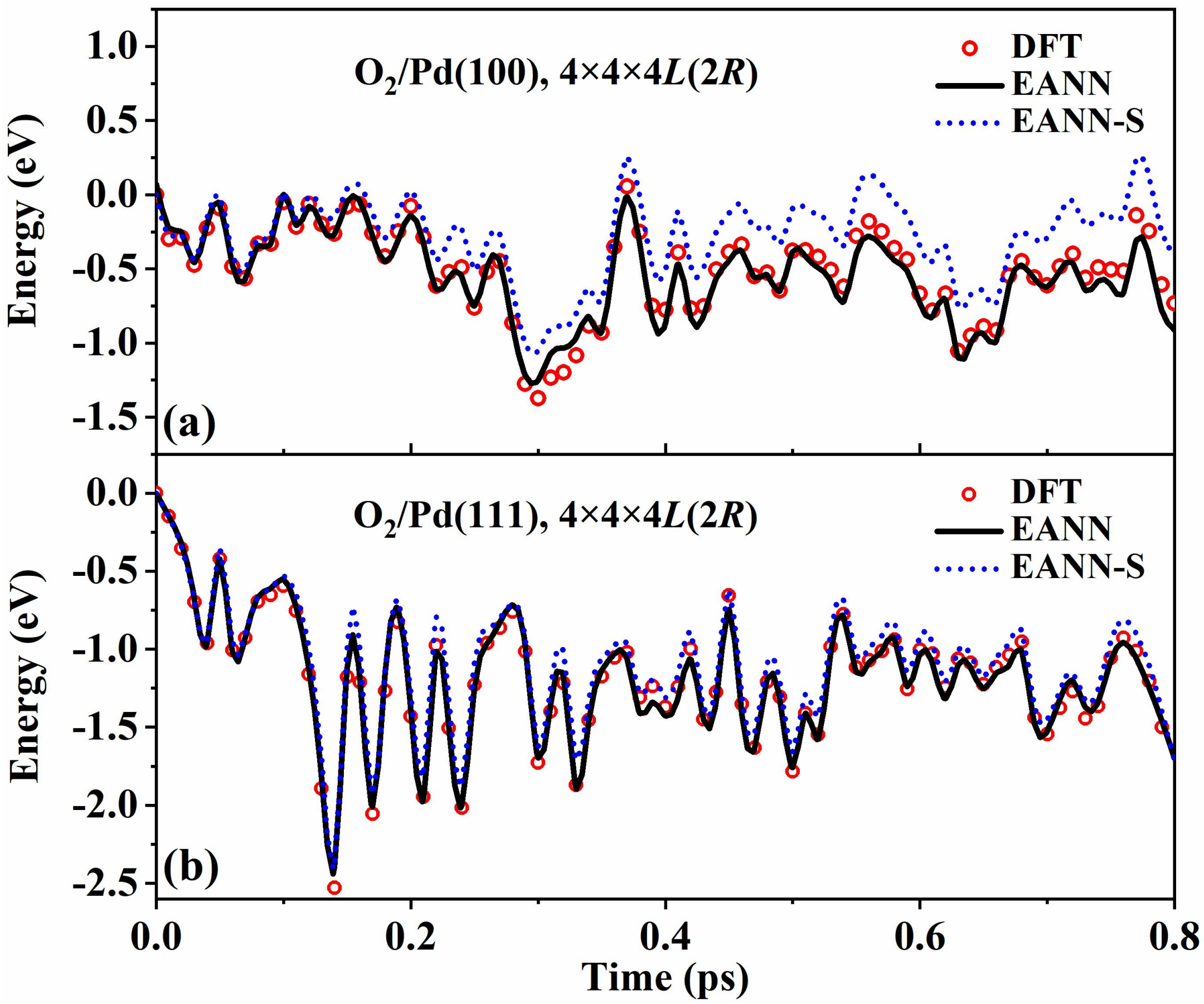

Figure S4. Comparison of the energy variation during a classical MD trajectory of $O_2$ dissociation on (a) Pd(100) and (b) Pd(111) with 4×4×4*L*(2*R*) supercells at a surface temperature of 160 K obtained from the EANN PES and direct DFT calculations. Here DFT data were not included for training.

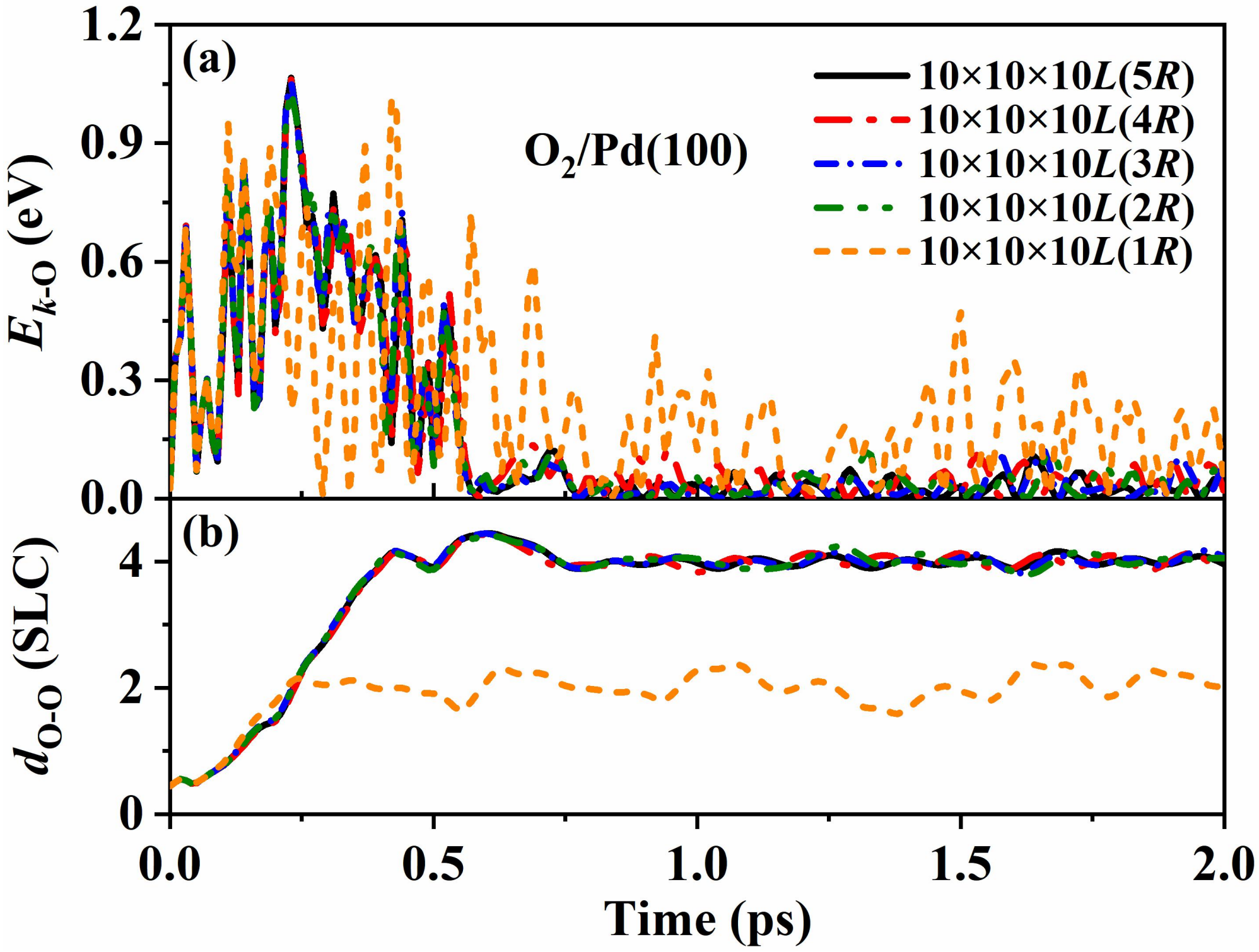

Figure S5. (a) Kinetic energy ($E_{k\text{-O}}$) and (b) the distance of oxygen atoms ($d_{\text{O-O}}$ per surface lattice constant, SLC) as a function of time along representative trajectories of $O_2$ dissociation on a 10×10×10*L* Pd(100) slab with different number of mobile Pd layers.

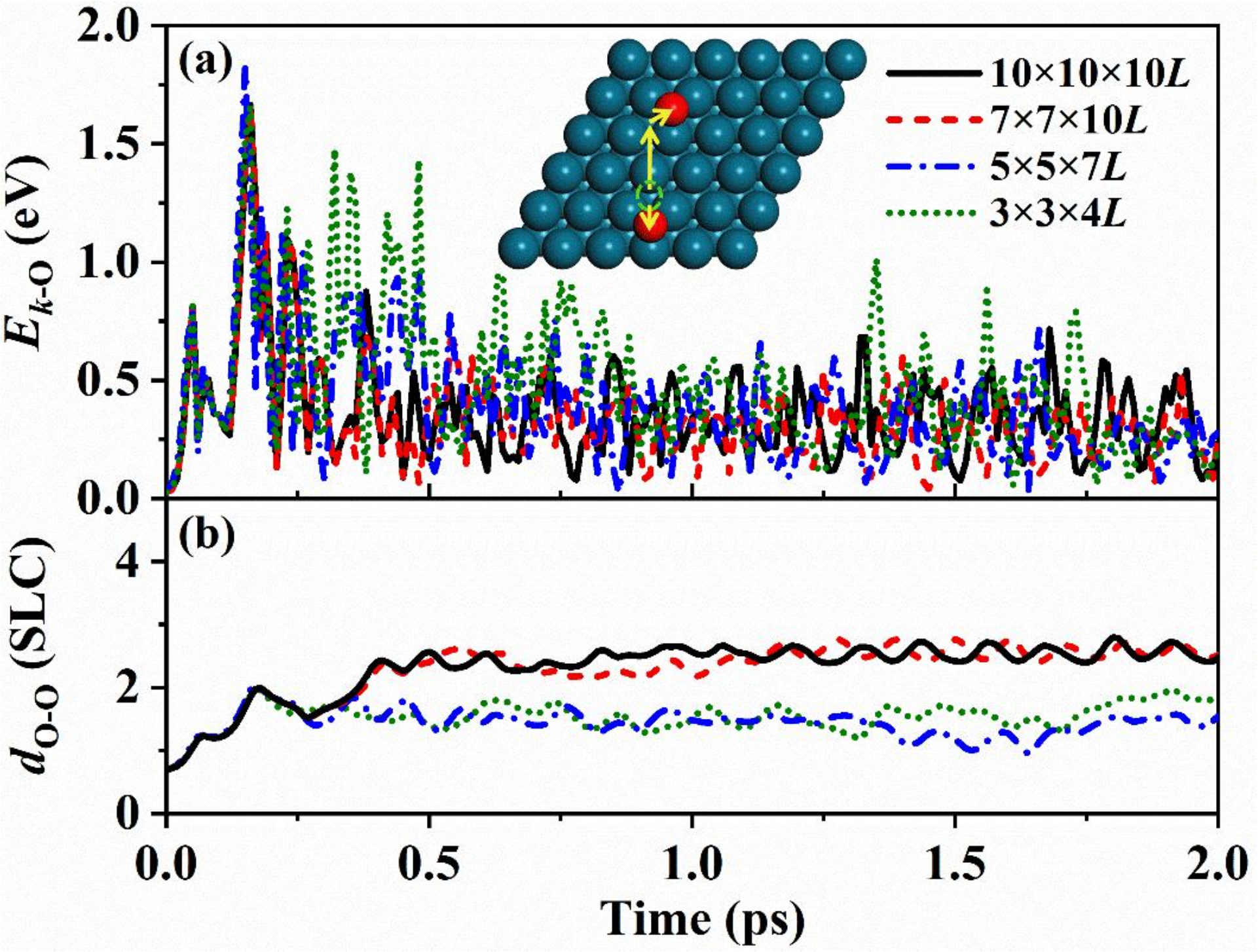

Figure S6. (a) Kinetic energy ($E_{k\text{-O}}$) and (b) the distance of oxygen atoms ($d_{\text{O-O}}$ per surface lattice constant, SLC) as a function of time along representative trajectories of $O_2$ dissociation on Pd(111) with different supercell sizes. The diffusion path is schematically represented in a top view (see inset).

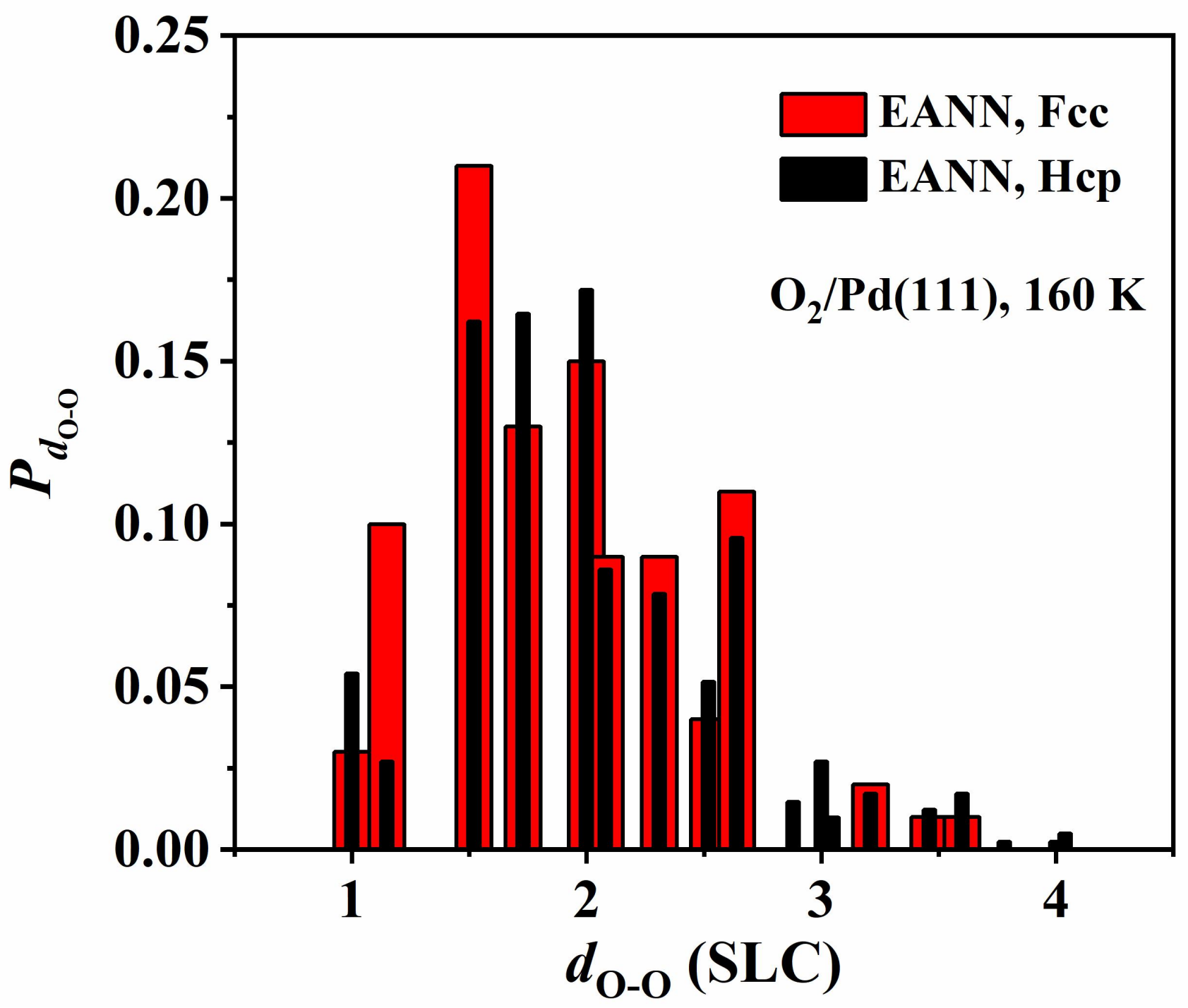

Figure S7. Comparison of the distance distributions of oxygen atom pairs after $O_2$ dissociation on Pd(111) initialized at fcc and hcp sites at a surface temperature of 160 K. See text for details.

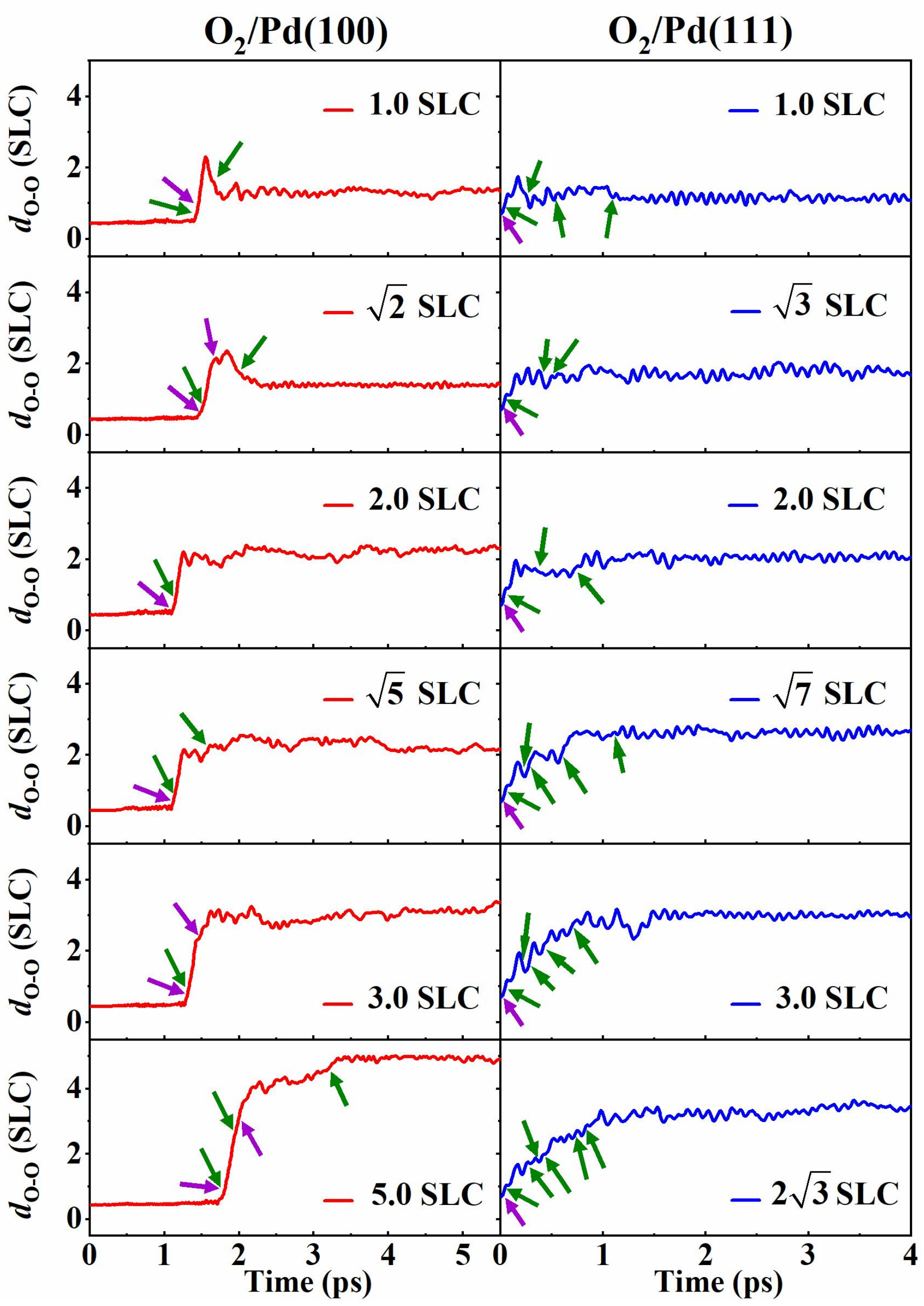

Figure S8. Distance of oxygen atoms ($d_{\text{O-O}}$) as a function of time along representative trajectories of $O_2$ dissociation on Pd(100) and Pd(111) at a surface temperature of 160 K with different equilibrium distances. The green and purple arrows represent each hop of two oxygen atoms.

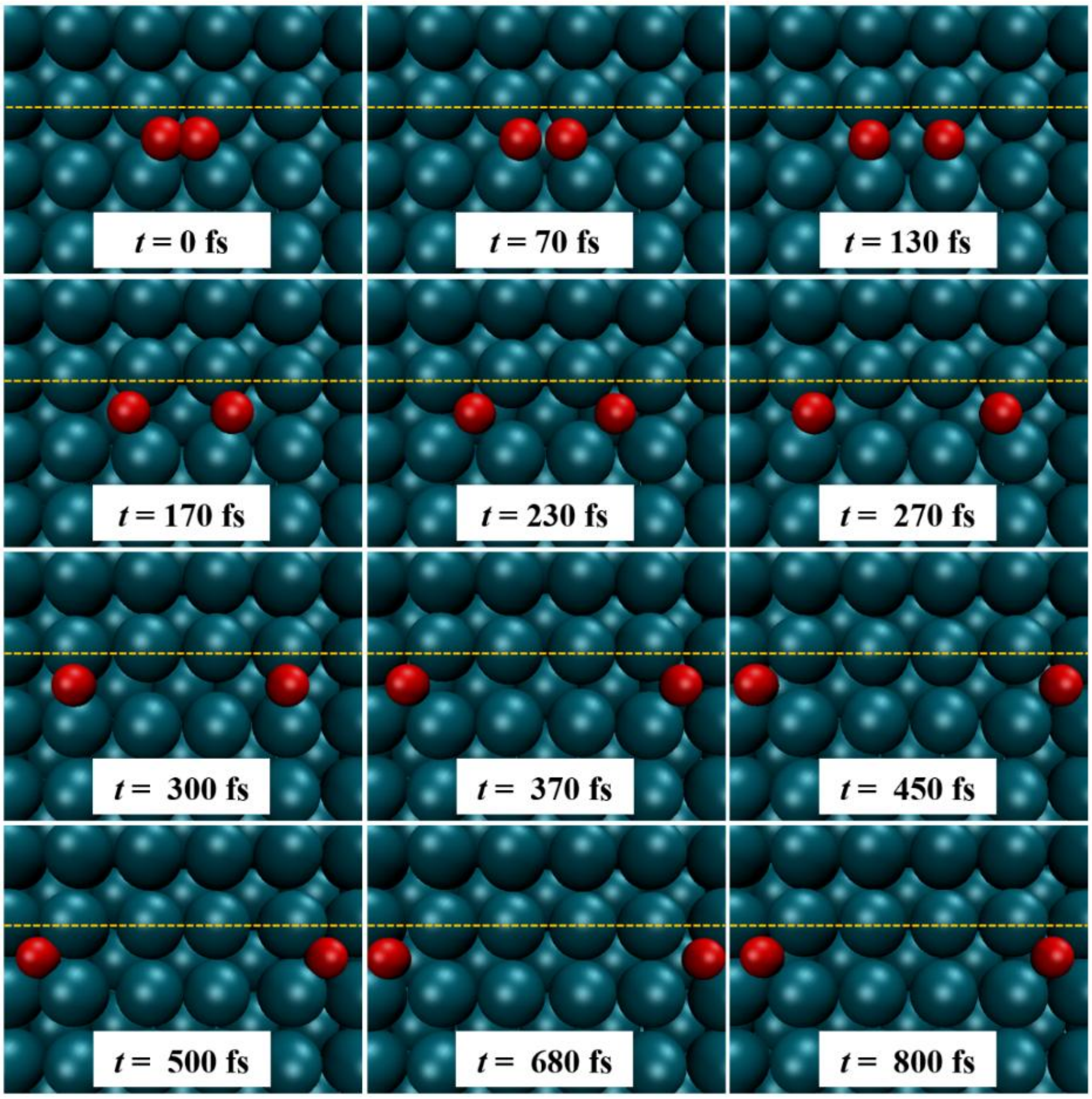

Figure S9. Evolution of atomic oxygen positions during the exemplary trajectory of $O_2$ dissociation on Pd(100) ending at $d_{O\text{-}O}$ of four times the surface lattice constant (SLC) (see Fig. 2). The yellow dashed lines are marks for equilibrium positions of Pd atoms.

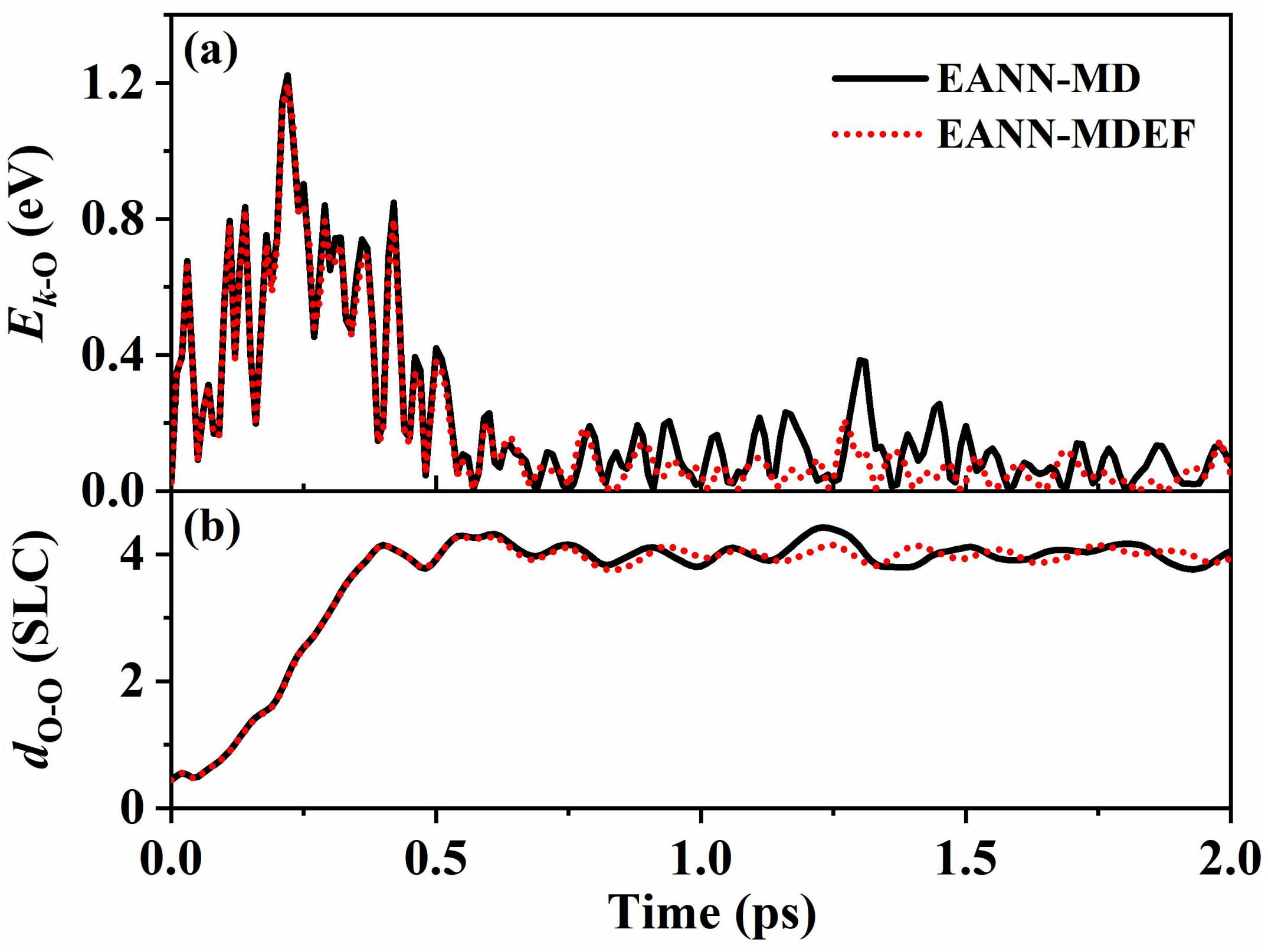

Figure S10. Comparison of (a) the kinetic energy ($E_{k\text{-O}}$) and (b) the distance of oxygen atoms ($d_{\text{O-O}}$) as a function of time along exemplary trajectories of $O_2$ dissociation on 3×8×4*L* Pd(100) slabs with (MD) and without (MDEF) electron friction.